\documentclass[a4paper, 12pt]{article}
\usepackage[T1, T2A]{fontenc}
\usepackage[english]{babel}
\usepackage{graphicx}
\usepackage{amssymb,amsfonts,amsmath,mathtext,cite,enumerate,float,microtype}
\usepackage {indentfirst}
\date{}

\title{UV Divergences of Scattering Amplitudes in D-dimensional Yang-Mills Theories}
\author{A.T. Borlakov and D.I. Kazakov\\[0.3cm]
Bogoliubov Laboratory of Theoretical Physics, \\ Joint Institute for Nuclear Research, Dubna, Russia}

\begin{document}

\maketitle

\newcommand{\nc}{\newcommand}
\nc{\beq}{\begin{equation}} \nc{\eeq}{\end{equation}}
\nc{\beqa}{\begin{eqnarray}} \nc{\eeqa}{\end{eqnarray}}
\nc{\ba}{\begin{array}} \nc{\ea}{\end{array}}
\nc{\R}{{\cal R}}
\nc{\A}{{\cal A}}
\nc{\K}{{\cal K}}
\nc{\B}{{\cal B}}

\abstract{We consider UV divergences for the on-shell planar gluon-gluon scattering amplitudes in the gauge theory in arbitrary D-dimensions in the one-loop order. The amplitudes are evaluated using the standard Feynman diagram technique for several choices of external gluon polarizations. Using the Passarino - Veltman reduction, the expansion in terms of the scalar master integrals is constructed and their calculation is performed within dimensional regularization. The resulting expressions are confronted with existing results for various cases, including maximally supersymmetric theories in  D=4, 6, 8, and 10 dimensions. One finds that, contrary to the D=4 case, in D=6 one-loop UV divergences cancel, while in D=8 and 10 the contributions of gauge, scalar and fermion loops have the same sign.}

\section{Introduction}

In recent years, considerable attention has been paid to the scattering amplitudes in supersymmetric gauge theories~\cite{Bartels, Dixon, BDS}. This is mostly due to the promising results obtained in the N=4 super Yang-Mills theories and hopes to get exact expressions for all loops and their connection with integrable models~\cite{Beisert}. In a series of papers~\cite{we0,we1,we2,we3,we4},
we considered the scattering amplitudes in gauge theories with maximal supersymmetry: D=6 N=2, D=8 N=1 and D=10 N=1 SYM theories. The choice of these models was motivated by remarkable simplicity and universality of perturbation expansion in these cases. Using the spinor-helicity and superspace formalisms,
it was found that the four-point colour-ordered scattering amplitudes in the planar limit are described by a universal set of master integrals containing mostly box-type diagrams. All the bubbles and triangles cancel in all orders of  perturbation theory. This representation immediately shows the absence of UV divergences in the D=4 N=4 SYM theory but allows their appearance starting from 3 loops in the D=6 N=2 case and starting from 1 loop in the D=8 N=1 and D=10 N=1 cases. The aim of our investigation~\cite{we0,we1,we2,we3,we4} was to evaluate the leading divergences in all loops and their summation via the generalised RG equations.  This goal was achieved by constructing recurrence relations that connect the leading divergences of subsequent orders of perturbation theory so that one can evaluate the leading divergences in a pure algebraic way starting from the lowest (one loop) order. These recurrence relations can then be converted into generalised RG equations that have an integro-differential form. Solving these equations analytically for a particular sequence of diagrams or otherwise numerically, one can study the high energy behaviour of the amplitudes. These solutions demonstrate either an exponential growth or decrease in the amplitudes depending on the channel in the D=6 case, or the presence of the infinite sequence  of poles in the D=8 and D=10 cases. 

In the present paper, we study a pure Yang-Mills theory in various dimensions. It is more complicated than the supersymmetric version since no cancellation takes place, all the diagrams including bubbles and triangles are present, and the number of master integrals increases. Since we know that the D=4 case possesses the asymptotic freedom regime, it is interesting to find out what happens for D > 4. Trying to use the spinor helicity formalism in gauge theories  in arbitrary D dimensions, we encountered difficulties in defining both internal and external polarization states in arbitrary D-dimensions. Therefore, we used the usual Feynman diagram technique and calculated the four-point amplitudes with various external polarizations. A particular choice of polarization does not affect the final result but allows one to get rid of vector indices and reduce the task to a set of scalar master integrals.
This set is universal for all dimensions but the coefficients of the master integrals, contrary to the supersymmetric case,  depend on particular dimension.

We limit ourselves here to the one loop case, construct master integrals and evaluate the coefficients. Then we analyse different limiting cases and check that adding fermions allows one to reproduce the known supersymmetric results. We leave the RG analysis for a subsequent publication.

\section{Preliminaries}

As already mentioned, to calculate the on shell scattering amplitude for the process $gg \rightarrow gg$, we use the standard Feynman diagram technique. This procedure requires the choice of polarization for external vector particles. For this purpose we first choose a reference frame, which can be arbitrary but is needed to fix the components of the polarization vectors. In what follows, we choose the center-of-mass frame (COM). If all external particles are treated as ingoing, their components for a massless theory  in the COM frame can be chosen as
\begin{equation}
\begin{aligned}
p_1&= (p, p , 0 , 0 , 0,...0),\\
p_2& = (p, -p , 0 , 0 , 0,...0),\\
p_3&= (-p, -p \, \cos{\theta}, -p \, \sin{\theta} , 0 , 0,...0),\\
p_4& =( -p, p \, \cos{\theta} , p \, \sin{\theta} , 0 , 0,...0),
\end{aligned}
\end{equation}
so that the Mandelstam variables take the form:

\begin{equation}
\begin{aligned}
	s &= (p_1+ p_2)^2 = 4 \, p^2,\\
	t &= (p_2 + p_3)^2 = -2 \, p^2(1+\cos{\theta}),\\
	u & = (p_1 + p_3)^2 = -2 \, p^2(1-\cos{\theta}).
\end{aligned}
\end{equation}
Then, expressing everything in terms of the Mandelstam variables, one has
\begin{equation}
\begin{aligned}
	p &= \sqrt{s/4},\\
	\cos{\theta} &= -\frac{s+2t}{s},\\
	\sin{\theta} &= \frac{2}{s}\sqrt{tu}.
\end{aligned}
\end{equation}

\subsection{Polarization vectors basis}
To define properly the polarization vectors in D-dimensions, we first consider the case of D=4.
In  general, the number of gluonic polarization states is equal to D-2, so we have two polarization vectors in D=4. We denote them as  <<+>> и <<->>. They have to satisfy the following conditions~\cite{Schwinn}: 
\begin{itemize}
	\item Transversality
	\begin{equation}
	\label{trans}
		p_\mu \epsilon_\pm^\mu(p) = 0 
	\end{equation}
	\item Normalization (here $\lambda=\pm$  is called helicity)	
	\begin{equation}
	\label{norm}
		\epsilon_\lambda(p) \cdot \epsilon_{\lambda'}^*(p) = -\delta_{\lambda,\lambda'}
	\end{equation}
	\item Completeness
	\begin{equation}
		\label{compl}
		\sum\limits_{\lambda=\pm} \epsilon_\lambda^\mu(p)\epsilon_\lambda^{\nu*}(p) = -\eta^{\mu\nu} + n^\mu \overline{n}^\nu + \overline{n}^\mu n^\nu,
	\end{equation}
\end{itemize}
where $\overline{n}^\mu$ and ${n}^\mu$ are the light-like reference momenta defined as (for example, for momentum $p_1$)
\begin{equation}
	\overline{n}^\mu=\frac{1}{\sqrt{2}}(1,1,0,0), \ \ \ {n}^\mu=\frac{1}{\sqrt{2}}(1,-1,0,0), \ \ \  n \cdot \overline{n} = 1, \ \ n^2=0, \ \ \overline{n}^2=0.
\end{equation}	
These two vectors together with the polarization ones form a complete basis in Minkowski space. Note that we can write
\begin{equation}
	\overline{n}^\mu=\frac{p^\mu}{(p\cdot n)}, 
\end{equation}
then the completeness relation becomes
\begin{equation}
\sum\limits_{\lambda=\pm} \epsilon_\lambda^\mu(p)\epsilon_\lambda^{\nu*}(p) = -\eta^{\mu\nu} + \frac{p^\mu n^\nu+ n^\mu p^\nu}{(p\cdot n)}.
\end{equation}
The conditions (\ref{trans})-(\ref{compl}) allow one to express the components of polarization vectors explicitly
\begin{equation}
 	\begin{aligned}
 	 \label{polvect4}
 		\epsilon_{+}(p_1)&=\frac{1}{\sqrt{2}}(0, 0, i, -1), &\epsilon_{-}(p_1)&=\frac{1}{\sqrt{2}}(0, 0, -i, -1),\\\epsilon_{+}(p_2)&=\frac{1}{\sqrt{2}}(0, 0, i, 1), &\epsilon_{-}(p_2)&=\frac{1}{\sqrt{2}}(0, 0, -i, 1),\\
		\epsilon_{+}(p_3)&=\frac{1}{\sqrt{2}}(0, -i \sin{\theta},i \cos{\theta},-1),
 			&		\epsilon_{-}(p_3)&=\frac{1}{\sqrt{2}}(0, i \sin{\theta},-i \cos{\theta},-1),\\
 			\epsilon_{+}(p_4)&=\frac{1}{\sqrt{2}}(0, -i \sin{\theta},i \cos{\theta},1),
 				&	\epsilon_{-}(p_4)&=\frac{1}{\sqrt{2}}(0, i \sin{\theta},-i \cos{\theta},1).
\end{aligned}
 \end{equation}
 	
Turning to the definition of polarization vectors in D = 6, we note that the conditions of transversality and completeness remain the same as in D = 4 (with $\mu=0,..,5$), and the normalization condition is given by\cite{Cheung}:
\begin{equation}
\begin{aligned}
\epsilon^{\mu}_{a\dot{a}}\epsilon_{\mu b\dot{b}}=\varepsilon_{ab}\varepsilon_{\dot{a}\dot{b}}
\end{aligned}
\end{equation}
where the indices $a$ and $\dot{a}$ running from 1 to 2 
belong to the little group
$SO(4) \simeq SU(2)\times SU(2)$.  The little group for $D$ dimensions is
$SO(D-2)$, so  for $D=4$ it is just $SO(2) \simeq U(1)$.   In the $D=6$ case
the action of the little group is no longer trivial and helicity is no longer conserved in contrast to the $D=4$ case. 

 Then, in the D=6 case, the polarization vectors for momentum $p_1$ can be taken in the form
\begin{equation}
 	\begin{aligned}
\epsilon_{1\dot{1}}(p_1)&=\frac{1}{\sqrt{2}}(0, 0, i, -1, 0, 0),\\
\epsilon_{2\dot{2}}(p_1)&=\frac{1}{\sqrt{2}}(0, 0, -i, -1, 0, 0),\\
\epsilon_{1\dot{2}}(p_1)&=\frac{1}{\sqrt{2}}(0, 0, 0, 0, i, -1),\\
\epsilon_{2\dot{1}}(p_1)&=\frac{1}{\sqrt{2}}(0, 0, 0, 0, i, 1).
	\end{aligned}
 \label{polvect6}
 	\end{equation}
 	
Note that the polarization vectors  $\epsilon_{1\dot{1}}(p_1),\epsilon_{2\dot{2}}(p_1)$
repeat the polarization vectors $\epsilon_{+}(p_1),\epsilon_{-}(p_1)$  in D=4 with two additional zero components at the end. 

Therefore, in what follows in D-dimensions we take the polarization vectors in the form of (\ref{polvect4})  with additional  D - 4 zero components and denote them as "+" and "-" like in four dimensions
\begin{equation}
 	\begin{aligned}
 	 \label{polvectD}
 		\epsilon_{+}(p_1)&=\frac{1}{\sqrt{2}}(0, 0, i, -1,\underbrace{0, ..., 0}_{D-4}), &\epsilon_{-}(p_1)&=\frac{1}{\sqrt{2}}(0, 0, -i, -1, \underbrace{0, ..., 0}_{D-4}),\\\epsilon_{+}(p_2)&=\frac{1}{\sqrt{2}}(0, 0, i, 1,\underbrace{0, ..., 0}_{D-4}), &\epsilon_{-}(p_2)&=\frac{1}{\sqrt{2}}(0, 0, -i, 1, \underbrace{0, ..., 0}_{D-4}),\\
		\epsilon_{+}(p_3)&=\frac{1}{\sqrt{2}}(0, -i \sin{\theta},i \cos{\theta},-1,\underbrace{0, ..., 0}_{D-4}),
 			&		\epsilon_{-}(p_3)&=\frac{1}{\sqrt{2}}(0, i \sin{\theta},-i \cos{\theta},-1,\underbrace{0, ..., 0}_{D-4}),\\
 			\epsilon_{+}(p_4)&=\frac{1}{\sqrt{2}}(0, -i \sin{\theta},i \cos{\theta},1,\underbrace{0, ..., 0}_{D-4}),
 				&	\epsilon_{-}(p_4)&=\frac{1}{\sqrt{2}}(0, i \sin{\theta},-i \cos{\theta},1,\underbrace{0, ..., 0}_{D-4}).
\end{aligned}
 \end{equation}

\subsection{Colour-ordering}

In order to calculate the amplitude, it is convenient to first extract the colour ordered partial amplitude by executing a colour decomposition~\cite{Smilga, Dixoncd}.
We start with the Lagrangian
\beq
{\cal L}= -\frac 14 Tr \hat F_{\mu\nu}\hat F_{\mu\nu}+\frac{1}{2\alpha}Tr \partial_\mu \hat A_\mu \partial_\nu \hat A_\nu +Tr \partial_\mu 
\hat c D_\mu \hat c,
\eeq
where the second term is the gauge fixing and the third is the ghost term; $\hat A_\mu=A^a_\mu T^a, \hat F_{\mu\nu}=\partial_\mu \hat A_\nu-\partial_\nu \hat A_\mu+ g[\hat A_\mu,\hat A_\nu]$ and the matrices $T^a$ obey the algebra
\beq
[T^a,T^b]=if^{abc}T^c.
\eeq
The structure constants $f^{abc}$ can be decomposed as $f^{abc} = -2i \, Tr[[T^a,T^b]T^c]$ and, using the Fierz identity
\begin{equation}
			T_j^{a,i}T_l^{a,k} = \frac{1}{2}(\delta_l^i\delta_j^k - \frac{1}{N_c}\delta_j^i\delta_l^k ),
			\label{cFierz}
\end{equation}
the product of the structure constants  can be written as 
\begin{equation}
	\begin{aligned}
	f^{abc}f^{cde} &= (-2i)^2Tr[[T^a,T^b]T^e]Tr[[T^c,T^d]T^e] 
	\\&= (-2i)^2[T^a,T^b]^i_j[T^c,T^d]^k_lT^{e,j}_iT^{e,l}_k\\
	&= -2 \, Tr[[T^a,T^b],[T^c,T^d]].
		\end{aligned}
\end{equation}
These relations allow one to present the n-particle amplitude in the colour ordered form
\begin{equation}
\mathcal{A}_n^{a_1\dots a_n,phys.}(p_1^{\lambda_1}\dots p_n^{\lambda_n})=\sum_{\sigma \in S_n/Z_n}Tr[\sigma(T^{a_1}\dots T^{a_n})]
\mathcal{A}_n(\sigma(p_1^{\lambda_1}\dots p_n^{\lambda_n}))+\mathcal{O}(1/N_c).
\end{equation}
The colour ordered amplitude $A_n$ is evaluated in the  limit  $N_c\to \infty$, $g^2\to 0$ and $g^2N_c$ is fixed, which corresponds to the planar diagrams.
In the case of the four-point amplitudes, the colour decomposition is performed as follows:
\begin{equation}
\mathcal{A}_4^{a_1\dots a_4,phys.}(1,2,3,4)=T^1\mathcal{A}_4(1,2,3,4)+T^2\mathcal{A}_4(1,2,4,3)+
T^3\mathcal{A}_4(1,4,2,3),
\end{equation}
where $T^i$ denote the trace combinations of $SU(N_c)$ generators in the fundamental representation 
\begin{gather}
T^1=Tr(T^{a_1}T^{a_2}T^{a_3}T^{a_4})+Tr(T^{a_1}T^{a_4}T^{a_3}T^{a_2}),\nonumber \\
T^2=Tr(T^{a_1}T^{a_2}T^{a_4}T^{a_3})+Tr(T^{a_1}T^{a_3}T^{a_4}T^{a_2}),\\
T^3=Tr(T^{a_1}T^{a_4}T^{a_2}T^{a_3})+Tr(T^{a_1}T^{a_3}T^{a_2}T^{a_4}).\nonumber
\end{gather}

Factorising the tree-level amplitude, which becomes a scalar factor after multiplying by the polarization vectors, the colour decomposed amplitude can be represented as
\beq
\mathcal{A}_4(1,2,3,4)=\mathcal{A}_4^{(0)}(1,2,3,4)M_4(1,2,3,4)=
\mathcal{A}_4^{(0)}(1,2,3,4)M_4([1+2]^2,[2+3]^2)\nonumber
\eeq
or using the standard Mandelstam variables
\beq
\mathcal{A}_4(1,2,3,4)=\mathcal{A}_4^{(0)}(1,2,3,4)M_4(s,t).
\eeq
The factorized colour ordered amplitude is expanded in PT series over the coupling $g^2$
\beq
M_4(s,t)=\sum_{L=1}^\infty (g^2)^L M_4^{(L)}(s,t).
\eeq
The one-loop factorised partial amplitude $M_4^{(1)}(s,t)$ is the subject of calculation in this paper. It can be expressed in terms of some universal combination of pure scalar master integrals times some coefficients which are the polynomials of the Mandelstam variables depending on D.

In what follows, we use the following Feynman rules for the colour-ordered amplitudes (in Feynman gauge)
\cite{Bern,Chen}:
\begin{figure}[ht]
\center{\includegraphics[scale=0.55]{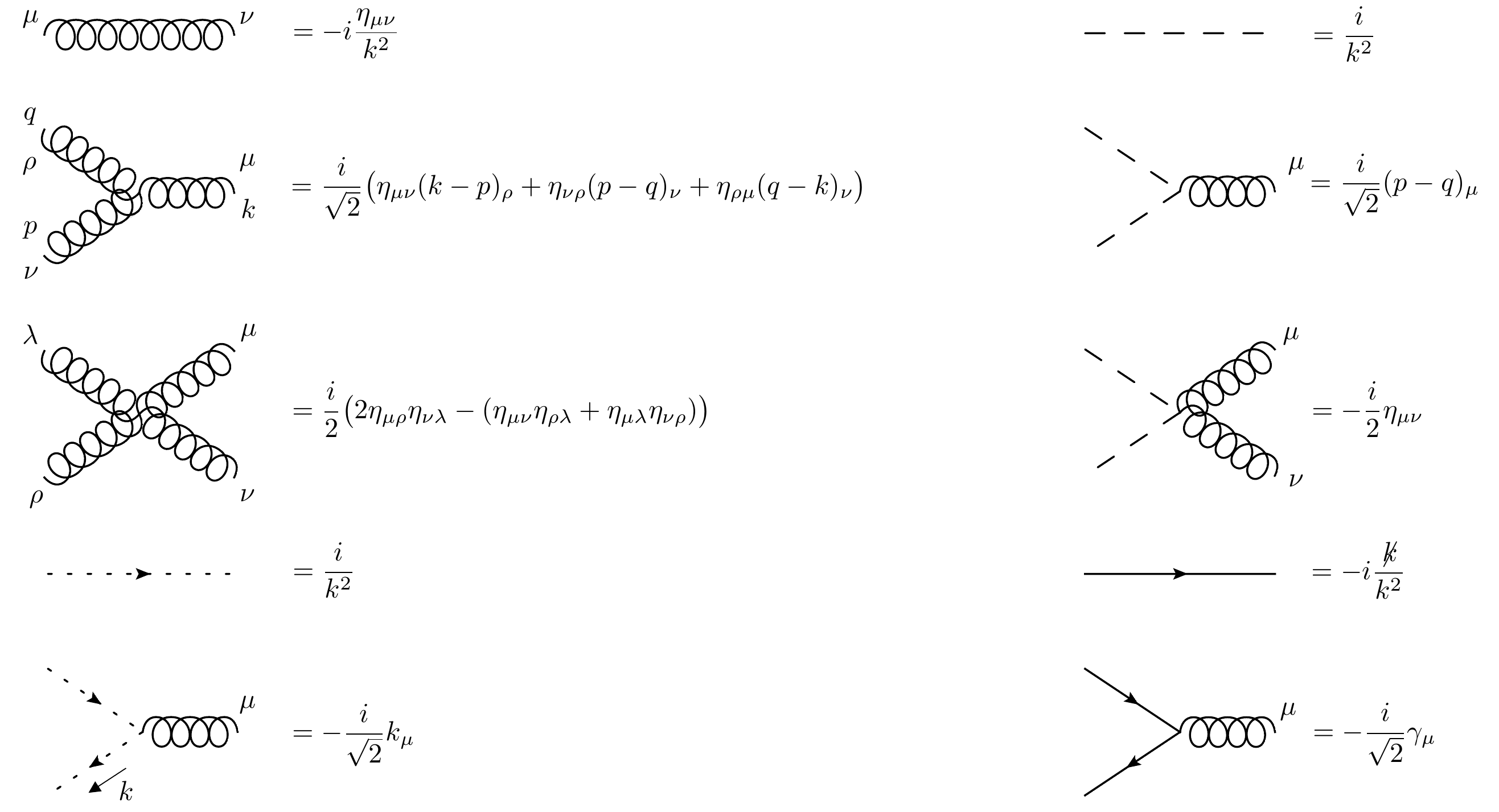}}
\caption{Colour-ordered Feynman rules in Feynman gauge }
(left column for gluons and ghosts  and right column for scalars and fermions)
\label{cofr}
\end{figure}

\section{Calculation of the Amplitudes}

\subsection{Tree Level}

Using the colour-ordered Feynman rules of Fig.\ref{cofr} and contracting external momenta with polarization vectors, given by (\ref{polvect4}), we obtain the following values for the tree-level amplitudes:
\begin{align}
\label{allp}
A_4^{(0)}&(1^+,2^+,3^+,4^+)=0,\\
\label{1m}
A_4^{(0)}&(1^+,2^+,3^+,4^-)=0,\\
\label{stree}
A_4^{(0)}&(1^+,2^+,3^-,4^-)=-ig^2\frac{s}{t},\\
\label{ttree}
A_4^{(0)}&(1^+,2^-,3^-,4^+)=-ig^2\frac{t}{s},\\
\label{sttree}
A_4^{(0)}&(1^+,2^-,3^+,4^-)=-ig^2\frac{(s+t)^2}{st}.
\end{align}

\begin{figure}[ht]
\center{\includegraphics[scale=0.3]{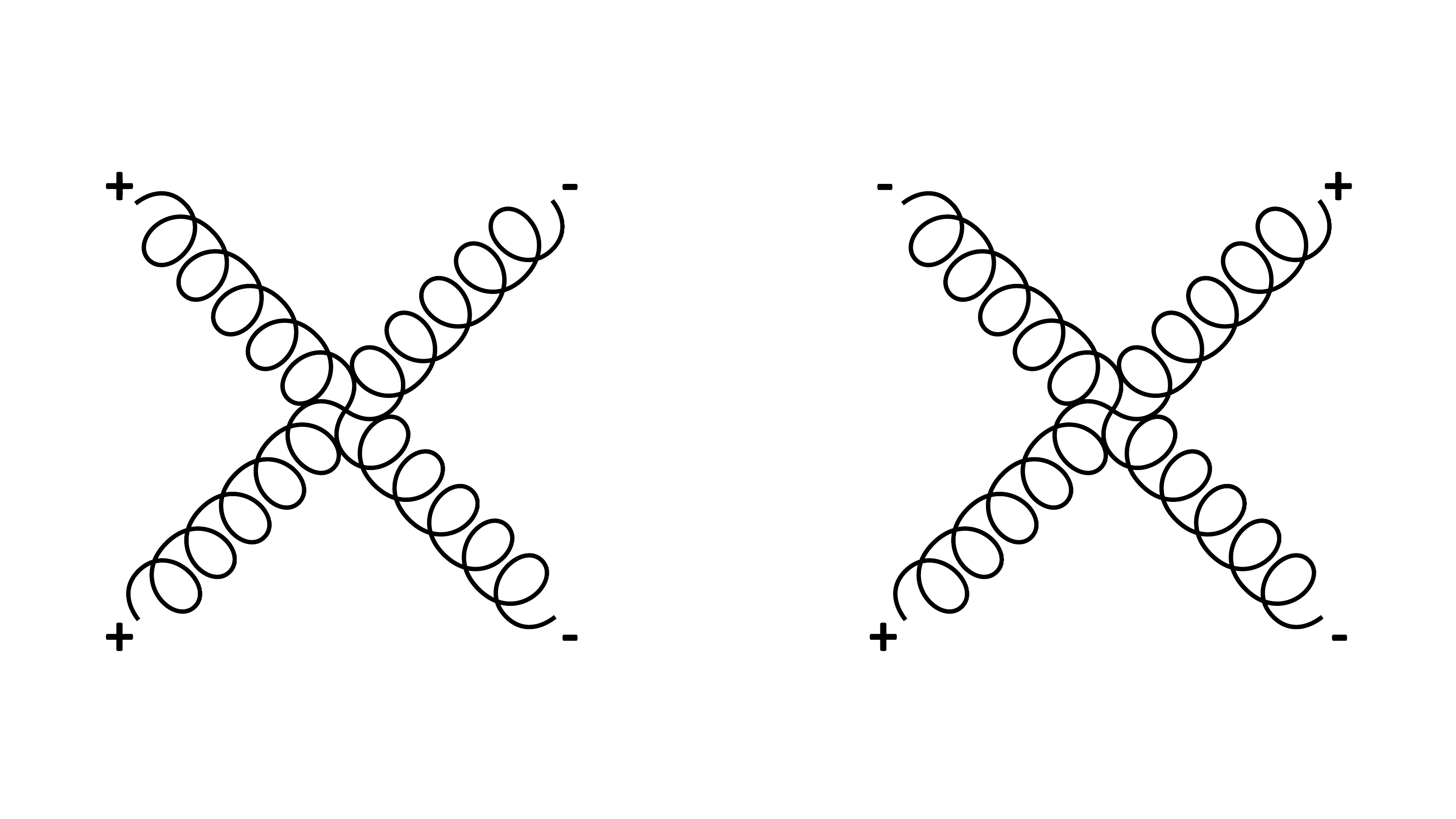}}
\caption{Adjacent and non-adjacent helicity cases}
\label{hel}
\end{figure}

As we see, the tree-amplitudes (\ref{allp}) и (\ref{1m}) are equal to zero, 
while the amplitudes (\ref{stree}) и (\ref{ttree}) are equal to each other up to the  substitution $s \to t , t \to s$. Hence, it is sufficient to consider the amplitudes with the following configurations of external particles, (+,+,--,--) and (+,--,+,--), namely the adjacent and non-adjacent case (see Fig.\ref{hel}).

\subsection{One-loop level}
\begin{figure}[ht]
\center{\includegraphics[scale=0.41]{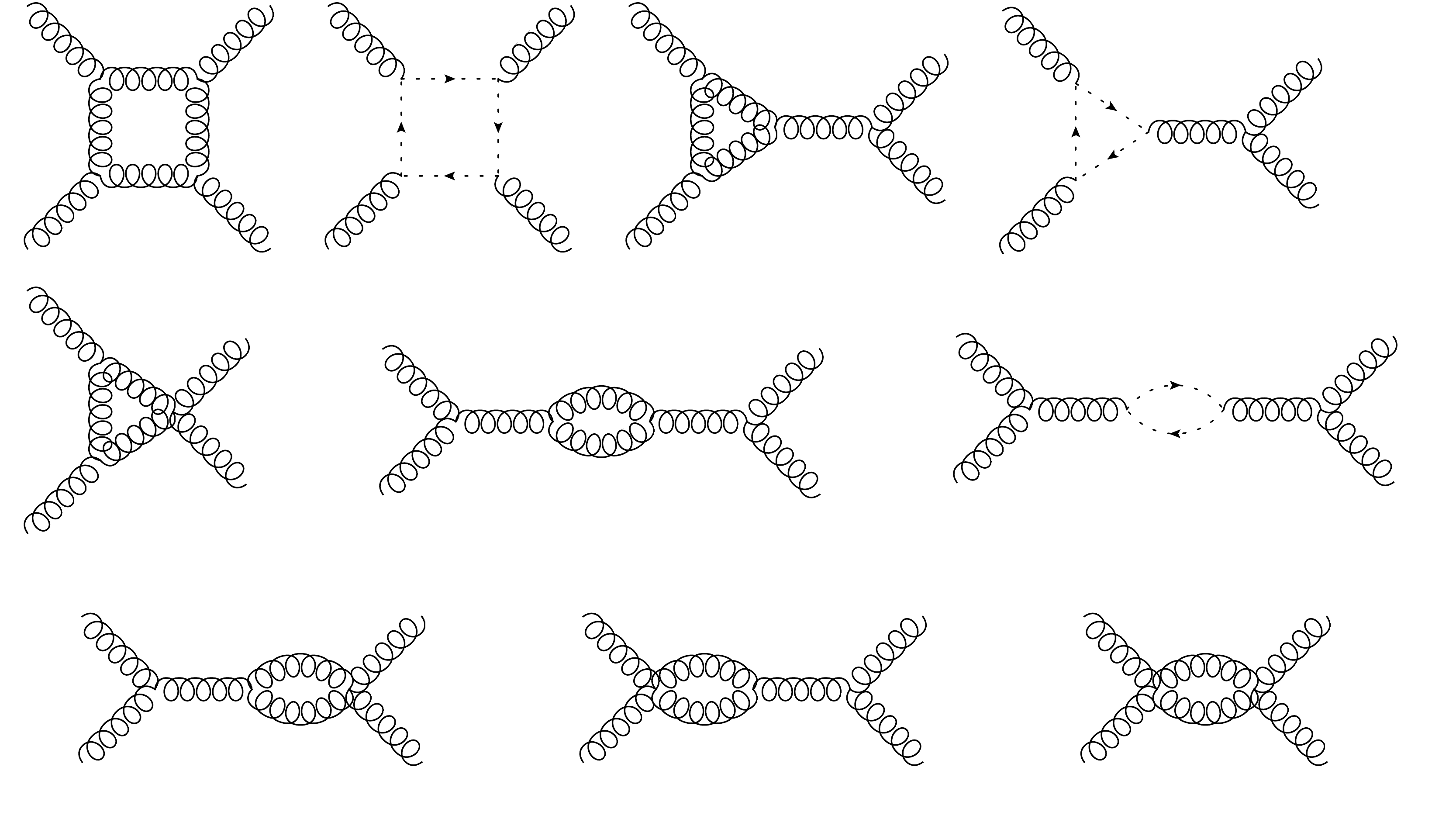}}
\caption{One-loop diagrams with gluons and ghosts in the loop}
\label{1loop}
\end{figure}

\begin{figure}[ht]
\center{\includegraphics[scale=0.42]{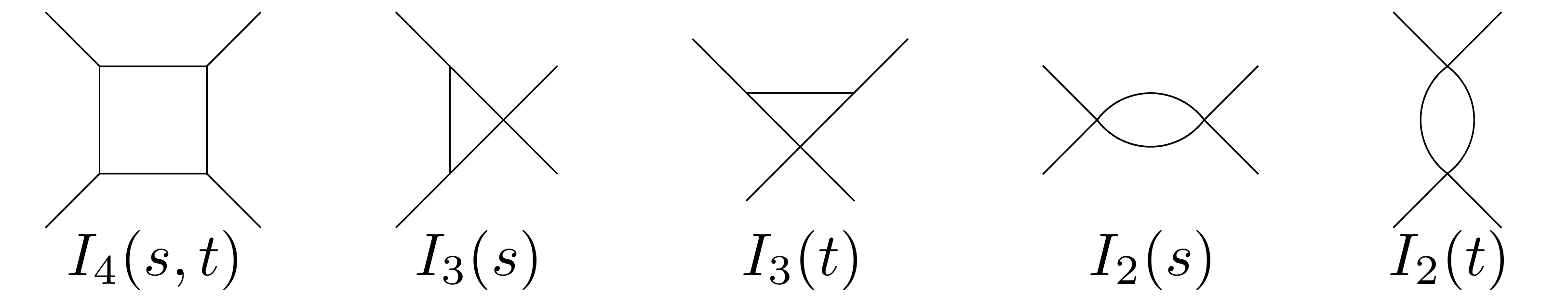}}
\caption{The set of scalar master integrals}
\label{MI}
\end{figure}

The one-loop contribution is given by the diagrams shown in Fig.\ref{1loop} (up to crossing symmetry). 
After contraction with polarization vectors, we arrive at a sum of expressions, where each of the terms is given by the integral
\begin{equation}
    T_n (p_1,...,p_{n-1},q)= \frac{1}{i \pi^2}\int d^D q \frac{Num}{D_1...D_{n-1}}
\end{equation}
with the denominator factors
\begin{equation}
    D_k = \left(q+\sum^{k-1}_{k=0}p_k \right)^2 +i\epsilon, \qquad k=0,...,n-1, \qquad p_0=0.
    \label{TInt}
\end{equation}
The numerator $Num$ depends on the scalar products $(q\cdot p_k)$ and on the tensor structures depending on $q^{\mu_1}...q^{\mu_n}$ which appeared after the contraction with the polarization vectors. Our aim is to represent the factorised colour-ordered  amplitude in the form
\begin{equation}
    M_4^{(1)} = C_4 I_4(s,t) + C_{3,1}I_3(s) + C_{3,2}I_3(t) + C_{2,1}I_2(s) + C_{2,2}I_2(t),
\end{equation}
where
\begin{equation}
    I_n (p_1,...,p_{n-1},q)= \frac{1}{i \pi^2}\int d^D q \frac{1}{D_1...D_{n-1}}
    \label{MInt}
\end{equation}
is the set of scalar master integrals (see Fig.\ref{MI}). 

To determine the coefficients $C_{k,i}$, we use the Passarino-Veltman reduction procedure~\cite{tHooft,Denner}. This procedure is implemented as part of the FeynCalc Wolfram Mathematica Package~\cite{FeynCalc1, FeynCalc2, FeynCalc3}, which we use in our calculations. Performing calculations in an arbitrary dimension, we distinguish the dependence that comes from the Lorentz contractions of the metric tensor  $\eta^\mu_\mu$ (we denote it as d) and the one coming from the measure of the Passarino-Veltman integrals (we denote it as D). This will allow us to consider various limiting cases later.

As a result, we obtain the following expressions for the massless gluon-gluon scattering amplitude both in the adjacent and non-adjacent cases
\beqa
&&\hspace{-1cm}M_4^{(1)}(1^+,2^+,3^-,4^-)= \\
&&\hspace{-1cm}-\frac{s t (16 (3\! - \!4 D \!+\! D^2) s^2\! + \!
   16 (2\! -\! 3 D\! +\! D^2) s t\! -\! (2 (16\! - \!14 D \!+ \!D^2)\! -\! 
      d (8\! -\! 6 D \!+ \!D^2)) t^2)}{16 (-3 + D) (-1 + D) (s + t)^2}I_4(s,t)\nonumber\\ &&\hspace{-1cm}-\frac{(-4 + D) s t (16 (-1 + D) s + (-20 - d (-2 + D) + 18 D) t)}{8 (-3 + D) (-1 + D) (s + t)^2}I_3(s)\nonumber\\ &&\hspace{-1cm}+\frac{(-4 + D) t^2 (-16 (-1 + D) s + (20 + d (-2 + D) - 18 D) t)}{8 (-3 + D) (-1 + D) (s + t)^2}I_3(t)\nonumber\\ &&\hspace{-1cm}-\frac{(-2 + d) (12 - 7 D + D^2) t }{4 (-3 + D) (-1 + D) (s + t)}I_2(s)+\frac{(2 (6 + d - 8 D) s \!+\! (20 + d (-2 + D) - 18 D) t)}{4 (-1 + D) (s + t)}I_2(t).\nonumber
 \label{nsym}
 \eeqa
\beqa
&&\hspace{-1cm}M_4^{(1)}(1^+,2^-,3^+,4^-)=\\
&&\hspace{-1cm}-\frac{1}{16 (-1 + D) (s + t)^4}\Big(\frac{1}{-3+D}s t (16 (3 - 4 D + D^2) s^4 + 16 (10 - 13 D + 3 D^2) s^3 t\nonumber \\&& \hspace{-1cm}+(224\! -\! 2 (146\! -\! d) D \!+\! (62\! +\! d) D^2) s^2 t^2\! +\! 
 16 (10\! - \!13 D\! +\! 3 D^2) s t^3 + 16 (3 \!- \!4 D\! +\! D^2) t^4)I_4(s,t)\nonumber \\ &&\hspace{-1cm}+\frac{1}{D-3}2 s^2 t (16 (2\! -\! 3 D\! +\! D^2) s^2\! +\! (64\! -\! 
      2 (46\! + \!d) D\! +\! (34\! -\! d) D^2) s t\! +\! 16 (2\! -\! 3 D\! +\! D^2) t^2) I_3(s)\nonumber \\ &&\hspace{-1cm}+\frac{1}{D-3}2 s t^2 (16 (2\! -\! 3 D\! +\! D^2) s^2\! +\! (64 \!- \!
      2 (46\! +\! d) D\! +\! (34\! -\! d) D^2) s t\! +\! 16 (2\! -\! 3 D\! +\! D^2) t^2) I_3(t)\nonumber \\ &&\hspace{-1cm}+\frac{1}{-2 + D}4 t (s + t) (8 (2 + d - 6 D + 2 D^2) s^2 - (-2 + D) (20 - 34 D +       d (6 + D)) s t \nonumber \\ &&\hspace{-1cm}- 2 (6 + d - 8 D) (-2 + D) t^2)I_2(s)+\frac{1}{-2 + D}4 s (s + t) (-2 (6 + d - 8 D) (-2 + D) s^2\nonumber  \\ &&\hspace{-1cm}- (-2 + D) (20 - 34 D + 
     d (6 + D)) s t+8 (2 + d - 6 D + 2 D^2) t^2)I_2(t)\Big).\nonumber 
 \label{sym}
\eeqa

As we can see, we have non-symmetric polynomial coefficients in the adjacent case and symmetric in the non-adjacent, respectively. We will analyse the obtained results in several limiting cases below.

\section{Analysis of the results}

Consider some particular cases of formulas (\ref{nsym},\ref{sym}) corresponding to different choices of the parameters $d$ and $D$. A similar analysis from a different point of view was made in \cite{Boels2017}.

\subsection{$d\to$ arbitrary, $D\to4$} 

This limit corresponds to the d-dimensional Lorentz algebra and 4-dimensional integration in the loops.

To begin with, we take the case of $d=4, D=4$ which gives us a possibility to confront our results with the well known ones. Substituting these values to eq.(\ref{nsym},\ref{sym}), we get
\begin{itemize}
	\item Adjacent case
\begin{equation}
    C_4= -s t,\ \  C_{3,1}=0, \ \ C_{3,2}=0, \ \ C_{2,1}=0, C_{2,2}=-\frac{11}{3}
\end{equation}	
	\item Non-adjacent case	
\end{itemize}
\begin{eqnarray}
	 && C_4=-\frac{s t (s^2 + s t + t^2)^2}{(s + t)^4},\\  &&C_{3,1}=-\frac{2 s^2 t (2 s^2 + 3 s t + 2 t^2)}{(s + t)^4},\ \  C_{3,2}=-\frac{2 s t^2 (2 s^2 + 3 s t + 2 t^2)}{(s + t)^4},\nonumber \\ &&C_{2,1}=-\frac{t (14 s^3 + 33 s^2 t + 30 s t^2 + 11 t^3) }{3 (s + t)^4}, C_{2,2}=-\frac{s (11 s^3 + 30 s^2 t + 33 s t^2 + 14 t^3)}{3 (s + t)^4}. \nonumber  
\end{eqnarray}

These results are in full agreement with those obtained with the help of the generalised-unitarity methods \cite{GU1,GU2}. To get the amplitude, one has to substitute the corresponding UV-divergent parts of the integrals in $D=4-2\epsilon$ 
\begin{equation*}
    Sing\ I_4(s,t) = 0, \ Sing\ I_3(s) = 0, \ Sing\ I_3(t) = 0, \ Sing I_2(s) = \frac 1\epsilon, \ Sing\ I_2(t) = \frac 1\epsilon
\end{equation*}
to obtain the following expressions
\begin{equation}
M_4^{(1)}(1^+,2^+,3^-,4^-)|_{d=4,D=4} = -\frac{11}{3}\frac 1\epsilon,     
\end{equation}
\begin{equation}
M_4^{(1)}(1^+,2^-,3^+,4^-)|_{d=4,D=4} = -\frac{11}{3}\frac 1\epsilon,    
\end{equation}
for the amplitudes in both helicity cases. This is the famous result for the $\beta_0$ coefficient for the running coupling in the 4-dimensional  Yang-Mills theory\cite{Gross}.

Now let us consider the $d=6, D=4$ case. The coefficients are given by
\begin{itemize}
	\item Adjacent case
\begin{equation}
    C_4= -s t, \ \ C_{3,1}=0, \ \ C_{3,2}=0, \ \ C_{2,1}=0, C_{2,2}=-\frac{10}{3}
\end{equation}	
	\item Non-adjacent case	
\beqa
	  &&\hspace{-0.5cm}C_4=-\frac{s t (s^4 + 2 s^3 t + 4 s^2 t^2 + 2 s t^3 + t^4) }{(s + t)^4},\\
	  &&\hspace{-0.5cm}C_{3,1}=-\frac{4 s^2 t (s^2 + s t + t^2)}{(s + t)^4}, \ \ C_{3,2}=-\frac{4 s t^2 (s^2 + s t + t^2)}{(s + t)^4},\nonumber\\ &&\hspace{-0.5cm}C_{2,1}=-\frac{2 t (8 s^3 + 15 s^2 t + 12 s t^2 + 5 t^3)}{3 (s + t)^4}, \ C_{2,2}=-\frac{2 s (5 s^3 + 12 s^2 t + 15 s t^2 + 8 t^3) }{3 (s + t)^4}. \nonumber  
\eeqa
\end{itemize}
A similar result for the adjacent case, which was obtained with the help of the generalised unitarity in six dimensions, can be found in \cite{Davies}. For the amplitudes, one again has in both cases
\begin{equation}
M_4^{(1)}(1^+,2^+,3^-,4^-)|_{d=6,D=4} = -\frac{10}{3} \frac 1\epsilon,     
\end{equation}
\begin{equation}
M_4^{(1)}(1^+,2^-,3^+,4^-)|_{d=6,D=4} = -\frac{10}{3}\frac 1\epsilon  .  
\end{equation}

At last, let us see what happens in the case of an arbitrary d. The coefficients in this case take the following form:
\begin{itemize}
	\item Adjacent case
\begin{equation}
    C_4= -s t, \ \ C_{3,1}=0, \ \ C_{3,2}=0, \ \ C_{2,1}=0, C_{2,2}=\frac{d-26}{6}
\end{equation}	
	\item Non-adjacent case	
	\begin{equation}
	\begin{aligned} 
	  C_4&=-\frac{s t (2 s^4 + 4 s^3 t + (2 + d) s^2 t^2 + 4 s t^3 + 2 t^4)) }{(s + t)^4},\\ C_{3,1}&=-\frac{s^2 t (4 s^2 - (-10 + d) s t + 4 t^2)}{(s + t)^4},\\ C_{3,2}&=-\frac{s t^2 (4 s^2 - (-10 + d) s t + 4 t^2) }{(s + t)^4},\\ C_{2,1}&=-\frac{t (2 (10 + d) s^2 + (58 - 5 d) s t - (-26 + d) t^2) }{6 (s + t)^4},\\ C_{2,2}&=\frac{s ((-26 + d) s^2 + (-58 + 5 d) s t - 2 (10 + d) t^2) }{6 (s + t)^4}  
	  \end{aligned} 
	\end{equation}
\end{itemize}
and the corresponding amplitudes are
\begin{equation}
M_4^{(1)}(1^+,2^+,3^-,4^-)|_{d,D=4} = \frac{d-26}{6} \frac 1\epsilon,
\label{adm26}
\end{equation}
\begin{equation}
M_4^{(1)}(1^+,2^-,3^+,4^-)|_{d,D=4} = \frac{d-26}{6} \frac 1\epsilon.
\label{nadm26}
\end{equation}
These expressions are universal. They do not depend on the choice of the polarization vector basis and the choice of gauge. The same expressions are also true for the off-shell UV-counter terms in d-dimensional Yang-Mills theories \cite{Kazakov2002}.

The same expressions can also be obtained in another way.
One can take the gluon contribution by performing the Lorentz algebra in d=4 and adding (d-4) scalar fields in the loops. The contribution of the scalar fields themselves are independent of d due to the absence of the Lorentz contractions in the gluon - scalar vertices. The scalar contribution is:
\begin{itemize}
	\item Adjacent case
\begin{equation}
    C_4= 0, \ \ C_{3,1}=0, \ \ C_{3,2}=0, \ \ C_{2,1}=0, \ \ C_{2,2}=\frac{1}{6}
\end{equation}

	\item Non-adjacent case	
	\begin{equation}
	\begin{aligned}
	  C_4&=-\frac{s^3 t^3 }{2(s + t)^4}, \ \ C_{3,1}=\frac{s^3 t^2 }{(s + t)^4}, \ \ C_{3,2}=-\frac{s^2 t^3}{(s + t)^4},\\ C_{2,1}&=\frac{t (-2 s^3 + 3 s^2 t + 6 s t^2 + t^3)}{6 (s + t)^4}, \ C_{2,2}=\frac{s (s^3 + 6 s^2 t + 3 s t^2 - 2 t^3) }{6 (s + t)^4}  
	  	\end{aligned}
	\end{equation}
\end{itemize}
Adding the scalar contribution, one gets
\beqa
&&M_4^{(1)}(1^+,2^+,3^-,4^-)|_{d=4,D=4}+(d-4)M_{4,scalar}^{(1)}(1^+,2^+,3^-,4^-)|_{D=4} = \frac{d-26}{6} \frac 1\epsilon, \nonumber \\    
&&M_4^{(1)}(1^+,2^-,3^+,4^-)|_{d=4,D=4}+(d-4)M_{4,scalar}^{(1)}(1^+,2^-,3^+,4^-)|_{D=4} = \frac{d-26}{6} \frac 1\epsilon, \nonumber
\eeqa
which are exactly the same as  (\ref{adm26}, \ref{nadm26}), as expected.

\subsection{$d\to D$, $D\to$ arbitrary}

Consider now the case when $d\to D$ for an arbitrary $D$. To be more concrete, we take $D$=6,\ 8 and 10.

For D=6 the coefficients of the master integrals are given by
\begin{itemize}
	\item Adjacent case
	\begin{equation}
	\begin{aligned} 
	  &C_4=-\frac{s t (240 s^2 + 320 s t + 112 t^2) }{240 (s + t)^2},\\ &C_{3,1}=-\frac{s t (80 s + 64 t)}{60 (s + t)^2},\\ &C_{3,2}=\frac{(-80 s - 64 t) t^2 }{60 (s + t)^2},\\ &C_{2,1}=-\frac{2 t}{5 (s + t)},\\ &C_{2,2}=\frac{(-72 s - 64 t)}{20 (s + t)}  
	  \end{aligned} 
	\end{equation}
	\item Non-adjacent case	
	\begin{equation}
	\begin{aligned} 
	  &C_4=-\frac{s t (240 s^4 + 640 s^3 t + 992 s^2 t^2 + 640 s t^3 + 240 t^4)}{240 (s + t)^4},\\ &C_{3,1}=-\frac{s^2 t (320 s^2 + 448 s t + 320 t^2)}{120 (s + t)^4},\\ &C_{3,2}=-\frac{s t^2 (320 s^2 + 448 s t + 320 t^2)}{120 (s + t)^4},\\ &C_{2,1}=-\frac{t (352 s^2 + 448 s t + 288 t^2)}{80 (s + t)^3},\\ &C_{2,2}=-\frac{s (288 s^2 + 448 s t + 352 t^2)}{80 (s + t)^3}  
	  \end{aligned} 
	\end{equation}
\end{itemize}
and the UV-divergent parts of the master integrals  in $D=6-2\epsilon$ are
\beqa
   && Sing\ I_4(s,t) = 0, \ Sing\ I_3(s) = -\frac {1}{2\epsilon}, \ Sing\ I_3(t) = -\frac {1}{2\epsilon},\nonumber \\ && Sing I_2(s) = \frac {s}{6\epsilon}, \ Sing\ I_2(t) = \frac {t}{6\epsilon}.
\eeqa
Adding all this together, we get the following result for the one-loop amplitude 
\begin{equation}
M_4^{(1)}(1^+,2^+,3^-,4^-)|_{d=6,D=6} = 0,      
\end{equation}
\begin{equation}
M_4^{(1)}(1^+,2^-,3^+,4^-)|_{d=6,D=6} = 0.     
\end{equation}
Thus,  in six dimensions the amplitude does not diverge in one loop. This is what happens in the supersymmetric case as well~\cite{we0,we1}.

For D=8, 10 the situation is different. Now the box diagram is also UV divergent. For D=8 the coefficients read
\begin{itemize}
	\item Adjacent case
	\begin{equation}
	\begin{aligned} 
	  &C_4=-\frac{s t (560 s^2 + 672 s t + 256 t^2) }{560 (s + t)^2},\\ &C_{3,1}=-\frac{s t (112 s + 76 t)}{70 (s + t)^2},\\ &C_{3,2}=\frac{(-112 s - 76 t) t^2 }{70 (s + t)^2},\\ &C_{2,1}=-\frac{6 t}{7 (s + t)},\\ &C_{2,2}=\frac{(-100 s - 76 t)}{28 (s + t)}  
	  \end{aligned} 
	\end{equation}
	\item Non-adjacent case	
	\begin{equation}
	\begin{aligned} 
	  &C_4=-\frac{s t (560 s^4 + 1568 s^3 t + 2496 s^2 t^2 + 1568 s t^3 + 560 t^4)}{560 (s + t)^4},\\ &C_{3,1}=-\frac{s^2 t (672 s^2 + 864 s t + 672 t^2)}{280 (s + t)^4},\\ &C_{3,2}=-\frac{s t^2 (672 s^2 + 864 s t + 672 t^2)}{280 (s + t)^4},\\ &C_{2,1}=-\frac{t (720 s^2 + 840 s t + 600 t^2)}{168 (s + t)^3},\\ &C_{2,2}=-\frac{s (600 s^2 + 840 s t + 720 t^2)}{168 (s + t)^3},  
	  \end{aligned} 
	\end{equation}
\end{itemize}
while the UV-divergent parts of the integrals in $D=8-2\epsilon$ are
\beqa
 &&   Sing\ I_4(s,t) = \frac {1}{6\epsilon}, \ Sing\ I_3(s) = -\frac {s}{24\epsilon}, \ Sing\ I_3(t) = -\frac {t}{24\epsilon}, \nonumber \\ && Sing I_2(s) = \frac {s^2}{60\epsilon}, \ Sing\ I_2(t) = \frac {t^2}{60\epsilon}.
\eeqa
Adding all together, one has for D=8
\begin{equation}
M_4^{(1)}(1^+,2^+,3^-,4^-)|_{d=8,D=8} = -\frac{4}{35\epsilon}st      \end{equation}
\begin{equation}
M_4^{(1)}(1^+,2^-,3^+,4^-)|_{d=8,D=8} = -\frac{29}{210\epsilon}st.    
\end{equation}
Notice different results for the adjacent and non-adjacent cases.

Finally, we consider the D=10 case. The coefficients of the master integrals are 
\begin{itemize}
	\item Adjacent case
	\begin{equation}
	\begin{aligned} 
	  &C_4=-\frac{s t (1008 s^2 + 1152 s t + 528 t^2)}{1008 (s + t)^2},\\ &C_{3,1}=-\frac{s t (144 s + 80 t)}{84 (s + t)^2},\\ &C_{3,2}=\frac{(-144 s - 80 t) t^2 }{84 (s + t)^2},\\ &C_{2,1}=-\frac{4 t}{3 (s + t)},\\ &C_{2,2}=\frac{(-128 s - 80 t)}{36 (s + t)}  
	  \end{aligned} 
	\end{equation}
	\item Non-adjacent case	
	\begin{equation}
	\begin{aligned} 
	  &C_4=-\frac{s t (1008 s^4 + 2880 s^3 t + 4704 s^2 t^2 + 2880 s t^3 + 1008 t^4)}{1008 (s + t)^4},\\ &C_{3,1}=-\frac{s^2 t (1152 s^2 + 1344 s t + 1152 t^2)}{504 (s + t)^4},\\ &C_{3,2}=-\frac{s t^2 (1152 s^2 + 1344 s t + 1152 t^2)}{504 (s + t)^4},\\ &C_{2,1}=-\frac{t (1216 s^2 + 1280 s t + 1024 t^2)}{288 (s + t)^3},\\ &C_{2,2}=-\frac{s (1024 s^2 + 1280 s t + 1216 t^2)}{288 (s + t)^3}, 
	  \end{aligned} 
	\end{equation}
\end{itemize}
while the UV-divergent parts of the master integrals in $D=10-2\epsilon$ are
\beqa
  &&  Sing\ I_4(s,t) = \frac {s+t}{120\epsilon}, \ Sing\ I_3(s) = -\frac {s^2}{360\epsilon}, \ Sing\ I_3(t) = -\frac {t^2}{360\epsilon}, \nonumber \\ && Sing I_2(s) = \frac {s^3}{840\epsilon}, \ Sing\ I_2(t) = \frac {t^3}{840\epsilon}.
\eeqa
Then for the D=10 amplitude one gets
\begin{equation}
M_4^{(1)}(1^+,2^+,3^-,4^-)|_{d=10,D=10} = -\frac{st(39s+49t)}{7560\epsilon} \end{equation}
\begin{equation}
M_4^{(1)}(1^+,2^-,3^+,4^-)|_{d=10,D=10} = -\frac{53st(s+t)}{7560\epsilon}. \end{equation}
Once again one has different results for the adjacent and non-adjacent cases. This reflects the fact that the amplitude in the case of gluonic external states depends on polarization, in contrast to the scalar or supersymmetric ones, which have no polarization.

Finally, consider the case of arbitrary d and D using the general formulas (\ref{nsym},\ref{sym}). The value of $d>D$ corresponds to a contribution of additional $d-D$ scalar fields, as we have demonstrated in the case of $D=4$ above.

For D = 6 we have
\begin{equation}
M_4^{(1)}(1^+,2^+,3^-,4^-)|_{d,D=6} = 0,      
\end{equation}
\begin{equation}
M_4^{(1)}(1^+,2^-,3^+,4^-)|_{d,D=6} = 0.     
\end{equation}
This means that the one loop amplitude in $D=6$ has no UV divergences, triangles and bubbles cancel each other in both gauge and scalar contributions.

For D = 8 and 10 the amplitudes are given by
\begin{equation}
M_4^{(1)}(1^+,2^+,3^-,4^-)|_{d,D=8} = -\frac{d+40}{420\epsilon}st,  
\end{equation}
\begin{equation}
M_4^{(1)}(1^+,2^-,3^+,4^-)|_{d,D=8} = -\frac{d+166}{1260\epsilon}st,
\end{equation}
\begin{equation}
M_4^{(1)}(1^+,2^+,3^-,4^-)|_{d,D=10} = -\frac{st(3(d+16)s+(d+88)t)}{15120\epsilon}, 
\end{equation}
\begin{equation}
M_4^{(1)}(1^+,2^-,3^+,4^-)|_{d,D=10} = -\frac{(d+202)st(s+t)}{30420\epsilon}.
\end{equation}\\
Here we can see that with increasing value of d (which is equal to the addition of extra scalars in the loop), the coefficients  become larger. This is due to the fact that the contribution from the gluon amplitude and from the scalars have the same sign. This is different from the D=4 case, where one has a negative sign for gauge fields and a positive sign for scalars. 
So one cannot reduce or cancel the divergent contribution of gluons by adding any number of scalar fields. 

\subsection{Supersymmetry check}

It is instructive to confront expressions (\ref{nsym}),(\ref{sym}) with the supersymmetric case considered earlier~\cite{Tseytlin, Sagnotti, Dennen, Berkovits, Caron-Huot, Boels}. It is known that for the maximally supersymmetric Yang-Mills (MSYM) theories there is a full cancellation of triangle and bubble diagrams, and the coefficient of the box diagram is universal and is equal to $-st$. The off-shell finitness of the 6D, $\mathcal{N}=(1,1)$ was studied in \cite{Buchbinder1, Buchbinder2} by using the harmonic superspace approach and in \cite{Buchbinder3} with the help of the component formulation.

For this purpose, one has to add the corresponding number of fermions and scalars to complete the gauge fields to a supersymmetry multiplet. The particle content of the MSYM theories in various dimensions is presented in Table.\ref{tab:MSYM} below~
\begin{table}[H]
\begin{center}
\begin{tabular}{|c|c|c|c|c|}
\hline
D & $\mathcal{N}$ & $n_g$ & $n_f$ & $n_s$ \\
\hline
4 & 4 & 1 & 4 & 6 \\
\hline
6 & 2 & 1 & 2 & 4 \\
\hline
8 & 1 & 1 & 1 & 2 \\
\hline
10 & 1 & 1 & 1 & 0 \\
\hline
\end{tabular}
\end{center}
\caption{\label{tab:MSYM}Particle content of various MSYM theories}
\end{table} 
There is some peculiarity about fermions. 
One has to define properly the fermion contribution in d-dimensions. The dependence of d comes from the trace of the $\gamma$-matrices
\begin{equation}
   \gamma^\mu\gamma_\mu = d; \quad\quad\quad Tr\gamma^\mu\gamma^\nu = \eta^{\mu\nu}Tr1 = \eta^{\mu\nu} = 2^{d/2}
   \label{fermion}
\end{equation}

In the cases of $D=4,6,8$, one has to take Majorana fermions reducing the number of degrees of freedom by a factor of 2, which results in an additional factor 1/2 for the fermion loop. For $D=10$  one imposes simultaneously Majorana and Weyl constraints on fermions reducing the number of degrees of freedom by a factor of 4, which results in an additional factor of 1/4 for the fermion loop.

We first take the case of $d=D=4$.   The scalar amplitude in D = 4 for any helicity is given by
\begin{equation}
M_{4,scalar}^{(1)}|_{D=4} = \frac{1}{6\epsilon},
\label{4s}
\end{equation}
while for the fermion amplitude  one has 
\begin{equation}
M_{4,fermion}^{(1)}|_{D=4} =  \frac 12\frac{4}{3\epsilon}.  
\label{4fw}
\end{equation}
For a theory with $n_f$ Majorana fermions and $n_s$ scalars we have
\begin{equation}
    M_4^{(1)} = M_{4,gluon}^{(1)} + n_f M_{4,fermion}^{(1)} + n_s M_{4,scalar}^{(1)}.
\end{equation}
Then,  using (\ref{adm26}), (\ref{nadm26}) along with (\ref{4s}) and (\ref{4fw}) we get
\begin{equation}
M_4^{(1)}|_{D=4} = -\frac{11}{3\epsilon} + \frac{2}{3\epsilon}n_f + \frac{1}{6\epsilon}n_s    
\end{equation}
and substituting $n_f=4,n_s=6$ we finally arrive at
\begin{equation}
M_4^{(1)}|_{D=4}^{UVdiv} = 0   
\end{equation}
as expected.

Turning to D-dimensions and having in mind (\ref{fermion}), the fermion amplitudes take the following form in the adjacent and non-adjacent cases, respectively
\begin{equation}
\begin{aligned}
&M_{4,fermion}^{(1)}(1^+,2^+,3^-,4^-) = \frac{2^{(d/2-4)} (4 - D) s t^2 (2 (-1 + D) s + D t)}{(-3 + D) (-1 + D) (s + t)^2}I_4(s,t)\\&-\frac{2^{(d/2-3)} (4\! -\!D) s t (2 ( D-1) s + D t)}{(-3 + D) (-1+D) (s + t)^2}I_3(s)-\frac{2^{(d/2-3)} (4\! -\! D) t^2 (2 (D-1) s + D t)}{(-3 + D) (-1 + D) (s + t)^2}I_3(t)\\&+\frac{2^{( d/2-2)} (-4 + D) t }{(-1 + D) (s + t)}I_2(s)+\frac{2^{(d/2-2)} (2 (-2 + D) s + D t)}{(-1 + D) (s + t)}I_2(t),
\end{aligned}  
\end{equation}
\begin{equation}
\begin{aligned}
&M_{4,fermion}^{(1)}(1^+,2^-,3^+,4^-) = 
\\&-\frac{2^{(d/
  2-4)} s^2 t^2 (2 (2\! -\! 3 D\! +\! D^2) s^2 + (8\! -\! 14 D\! +\! 3 D^2) s t + 
   2 (2\! -\! 3 D \!+\! D^2) t^2)}{(-3 + D) (-1 + D) (s + t)^4}I_4(s,t)\\&+\frac{2^{( d/2-3)} s^2 t (2 (2\! -\! 3 D\! +\! D^2) s^2 + (8\! -\! 14 D \!+\! 3 D^2) s t + 
   2 (2\! -\! 3 D\! +\! D^2) t^2) }{(-3 + D) (-1 + D) (s + t)^4}I_3(s)\\&+\frac{2^{( d/2-3)} s t^2 (2 (2\! -\! 3 D\! +\! D^2) s^2 + (8 \!-\! 14 D\! +\! 3 D^2) s t + 
   2 (2\! -\! 3 D\! + \!D^2) t^2) }{(-3 + D) (-1 + D) (s + t)^4}I_3(t)\\&+\frac{2^{( d/2-2)} t (2 (6 - 3 D + D^2) s^2 + (20 - 16 D + 3 D^2) s t + 2 (-2 + D)^2 t^2) }{(-2 + D) (-1 + D) (s + t)^3}I_2(s)\\&+\frac{2^{(d/2-2)} s (2 (-2 + D)^2 s^2 + (20 - 16 D + 3 D^2) s t + 2 (6 - 3 D + D^2) t^2)}{(-2 + D) (-1 + D) (s + t)^3}I_2(t).
\end{aligned}  
\end{equation}

For the scalar contribution one has
\begin{equation}
\begin{aligned}
&M_{4,scalar}^{(1)}(1^+,2^+,3^-,4^-) = \frac{(-8 + 6 D - D^2) s t^3}{16 (-3 + D) (-1 + D) (s + t)^2}I_4(s,t)\\&+\frac{(8 - 6 D + D^2) s t^2}{8 (-3 + D) (-1 + D) (s + t)^2}I_3(s)+\frac{(8 - 6 D + D^2) t^3}{8 (-3 + D) (-1 + D) (s + t)^2}I_3(t)\\&-\frac{(12 - 7 D + D^2) t}{4 (-3 + D) (-1 + D) (s + t)}I_2(s)+\frac{(2 s + (-2 + D) t) }{4 (-1 + D) (s + t)}I_2(t)
\end{aligned}  
\end{equation}
\begin{equation}
\begin{aligned}
&M_{4,scalar}^{(1)}(1^+,2^-,3^+,4^-) = -\frac{D (-4 + D^2) s^3 t^3}{16 (-3 + D) (-2 + D) (-1 + D) (s + t)^4}I_4(s,t)\\&+\frac{D (-4 + D^2) s^3 t^2}{8 (-3 + D) (-2 + D) (-1 + D) (s + t)^4}I_3(s)\\&+\frac{D (-4 + D^2) s^2 t^3}{8 (-3 + D) (-2 + D) (-1 + D) (s + t)^4}I_3(t)\\&-\frac{t (8 s^2 - (-12 + 4 D + D^2) s t - 2 (-2 + D) t^2) }{4 (-2 + D) (-1 + D) (s + t)^3}I_2(s)\\&+\frac{s (2 (-2 + D) s^2 + (-12 + 4 D + D^2) s t - 8 t^2)}{4 (-2 + D) (-1 + D) (s + t)^3}I_2(t).
\end{aligned}  
\end{equation}

Using these expressions (taking into account an additional factor of 1/2 for D=6, 8 and 1/4 for D=10 for the fermion contribution), along with (\ref{nsym}), (\ref{sym}), and with the particle content from Table.\ref{tab:MSYM},  one can check that for any helicity 
 \begin{equation}
M_4^{(1)}|_{D=4, 6, 8, 10} = -s t I_4(s,t),    
 \end{equation}
in agreement with the  MSYM case, thus verifying  our calculations.

\section{Conclusions}

The obtained formulas for the UV divergent contributions for one-loop planar gluon-gluon scattering amplitudes have a universal form and can be applied in any dimension. We distinguished the Lorentz algebra dimension d and the loop momenta dimension D mostly for the purpose of verification of the validity of our results and comparison with the known ones. To explore the purely D-dimensional case, one should put d=D in our formulas.

We concentrate on two aspects. The first one is the structure of the scalar master integrals. In the case of interest, one has all possible forms of diagrams  (bubbles, triangles and boxes) following from the Feynman rules. Contrary to the supersymmetric case, there are no cancellations of bubbles and triangles. This means that in two and more loops one can also expect a full set that can be reproduced using the recurrence relations, as was done in the theories we considered earlier~\cite{we2}. 

The second aspect is related to the sign of various contributions. As is well known, in D=4 the sign of the gauge contrubution is negative while the sign of the scalar and fermion ones is positive. The situation is different for D>4. One finds that the one-loop amplitude in D=6 has no UV divergences for both gauge and matter fields, while in D=8, 10, etc the signs of all the contributions are the same. It is basically negative but contains a polynomial dependence on the Mandelstam variables and can have different signs depending on the kinematics.

The last statement is important when trying to find the high energy behaviour of the amplitude.  The UV divergent terms $\sim 1/\epsilon$ are in one-to-one correspondence with the leading logarithms and one can sum them up using the generalized  RG equations. We demonstrated how this can be done for several QFT models~\cite{we4} and are going to repeat this procedure for the D-dimensional Yang-Mills theory in a subsequent publication.

\section*{Acknowledgements}
The authors are grateful to L.V.Bork and E.A.Ivanov for valuable discussions.
Financial support from the Russian Science Foundation, grant \#21-12-00129, is acknowledged.

\end{document}